\documentclass[10pt]{article}
\usepackage[latin1]{inputenc}
\usepackage[english]{babel}
\usepackage[none]{hyphenat}
\usepackage[T1]{fontenc}
\usepackage[left=2cm,right=2cm,top=1.5cm,bottom=2cm]{geometry}
\usepackage[usenames,dvipsnames]{color}  
\usepackage[pdfborder={0 0 0},colorlinks=true,urlcolor=blue,linkcolor=blue,citecolor=blue,bookmarks=true]{hyperref}
\usepackage{titlesec}
\usepackage[numbers]{natbib}
\usepackage{graphicx}
\usepackage[frozencache,cachedir=_minted-paper]{minted}
\usepackage{authblk}
\usepackage{amsmath}
\usepackage{amsthm}
\usepackage{bm}
\usepackage{booktabs}      
\usepackage{multirow}      
\usepackage{tcolorbox}   
\usepackage{times}
\usepackage{inconsolata}   
\usepackage{adjustbox}
\usepackage{threeparttable}
\usepackage{parcolumns}
\usepackage{amsfonts}
\usepackage{nicefrac}
\usepackage{microtype}
\usepackage{tabularx}
\usepackage{array}
\usepackage{wasysym}

\usepackage{microtype}
\DisableLigatures[f]{encoding = *, family = * }

\small\normalsize

\newcolumntype{R}[2]{%
  >{\adjustbox{angle=#1,lap=\width-(#2)}\bgroup}%
  l%
  <{\egroup}%
}
\newcommand*\rot{\multicolumn{1}{R{30}{2.0em}}}

\begin{document}

\pagestyle{empty}

\begin{center}

{\Large\bf Flexible numerical optimization with ensmallen}



\vspace{1.5ex}
Ryan R. Curtin{\tiny~}{$^{1,*}$}, 
Marcus Edel{\tiny~}{$^{2,3,*}$}, 
Rahul Ganesh Prabhu{\tiny~}{$^{4}$}, 
Suryoday Basak{\tiny~}{$^{5}$},
Zhihao Lou{\tiny~}{$^{6}$}, 
Conrad Sanderson{\tiny~}{$^{7,8}$}
\vspace{1.5ex}


\begin{small}
\begin{tabular}{l}
$^{1}$~RelationalAI, Atlanta, USA\\
$^{2}$~Free University of Berlin, Germany\\
$^{3}$~Collabora, Montreal, Canada\\
$^{4}$~Birla Institute of Technology and Science, Pilani, Pilani Campus, India\\
$^{5}$~Department of Computer Science and Engineering, The University of Texas at Arlington, USA\\
$^{6}$~Epsilon, Chicago, USA\\
$^{7}$~Data61~/~CSIRO, Australia\\
$^{8}$~Griffith University, Australia\\
$^{*}$~{\it Primary maintainers}
\end{tabular}
\end{small}

\end{center}

\section*{Abstract}

This report provides an introduction to the {\tt ensmallen} numerical optimization library,
as well as a deep dive into the technical details of how it works.
The library provides a fast and flexible C++ framework
for mathematical optimization of arbitrary user-supplied functions.
A large set of pre-built optimizers is provided,
including many variants of Stochastic Gradient Descent and Quasi-Newton optimizers.
Several types of objective functions are supported, including differentiable,
separable, constrained, and categorical objective functions.
Implementation of a new optimizer requires only one method,
while a new objective function requires typically only one or two C++ methods.
Through internal use of C++ template metaprogramming, {\tt ensmallen} provides support for arbitrary
user-supplied callbacks and automatic inference of unsupplied methods without
any runtime overhead.
Empirical comparisons show that {\tt ensmallen}
outperforms other optimization frameworks (such as Julia and SciPy), sometimes
by large margins.  The library is available at \url{https://ensmallen.org}
and is distributed under the permissive BSD license.


\vspace{4ex}
\hrule

\section{Introduction}
\label{sec:introduction}

The problem of mathematical optimization is fundamental in the computational
sciences~\cite{Nocedal_2006}.  In short, this problem is expressed as
\begin{equation}
\operatornamewithlimits{argmin}_x f(x)
\end{equation}

\noindent
where $f(x)$ is a given objective function and $x$ is typically a vector or matrix.
The ubiquity of this problem gives rise to the proliferation of mathematical
optimization toolkits, such as SciPy~\cite{2019arXiv190710121V},
opt++~\cite{meza1994opt++},
OR-Tools~\cite{ortools}, CVXOPT~\cite{vandenberghe2010cvxopt},
NLopt~\cite{johnson2014nlopt}, Ceres~\cite{ceres-solver},
and RBFOpt~\cite{costa2018rbfopt}.
Furthermore, in the field of machine learning, many
deep learning frameworks have integrated optimization
components.  Examples include Theano~\cite{2016arXiv160502688},
TensorFlow~\cite{tensorflow2015-whitepaper}, PyTorch~\cite{NEURIPS2019_9015},
and Caffe~\cite{jia2014caffe}.

Mathematical optimization is generally quite computationally intensive.
For instance, the training of deep neural networks is dominated by
the optimization of the model parameters on the
data~\cite{krizhevsky2012imagenet, lauzon2012introduction}.  Similarly,
other popular machine learning algorithms such as logistic regression are also
expressed as and dominated by an optimization process~\cite{zhang2004solving,
manogaran2018health}.  Computational bottlenecks occur even in fields as
wide-ranging as rocket landing guidance systems~\cite{dueri2016customized},
motivating the development and implementation of specialized solvers.

The necessity for both efficient and specializable function optimization motivated us
to implement the {\tt ensmallen} library, originally as part of the {\tt mlpack}
machine learning library~\cite{mlpack2018}.
The {\tt ensmallen} library provides a large {set of pre-built optimizers} for optimizing
{user-defined objective functions} in C++;
at the time of writing, 46 optimizers are available.  
The external interface to the optimizers is intuitive and matches the ease of use of popular
optimization toolkits mentioned above.

However, unlike many of the existing optimization toolkits,
{\tt ensmallen} explicitly supports numerous function types:
arbitrary, differentiable, separable, categorical, constrained, and semidefinite.
Furthermore, custom behavior during optimization can be easily specified via {\it callbacks}.
Lastly, the underlying framework facilitates the implementation of new optimization techniques,
which can be contributed upstream and incorporated into the library.
Table~\ref{tab:comparison} contrasts the functionality supported
by {\tt ensmallen} and other optimization toolkits.

This report is a revised and expanded version of our initial overview
of {\tt ensmallen}~\cite{ensmallen2018}. 
It also provides a deep dive into the internals
of how the library works, which can be a useful resource for anyone looking
to contribute to the library or get involved with its development.


We continue the report as follows.
We overview the available functionality and show example usage in Section~\ref{sec:overview}.
Advanced usage such as callbacks is discussed in Section~\ref{sec:advanced}.
We demonstrate the empirical efficiency of {\tt ensmallen} in Section~\ref{sec:experiments}.
The salient points are summarized in Section~\ref{sec:conclusion}.

\begin{table}[!t]
\centering
    \begin{tabular}{@{} cl*{9}c @{}}
        & & \multicolumn{7}{c}{} \\[0.6ex]
        & & \rot{unified framework}
          & \rot{constraints}
          & \rot{separable functions / batches}
          & \rot{arbitrary functions}
          & \rot{arbitrary optimizers}
          & \rot{sparse gradients}
          & \rot{categorical}
          & \rot{arbitrary types}
          & \rot{callbacks} \\
        \cmidrule[1pt]{2-11}
        & \texttt{ensmallen}            & \CIRCLE & \CIRCLE & \CIRCLE & \CIRCLE & \CIRCLE & \CIRCLE & \CIRCLE & \CIRCLE & \CIRCLE\\
        & Shogun \cite{sonnenburg2010shogun}             & \CIRCLE & - & \CIRCLE
& \CIRCLE & \CIRCLE & - & - & - & - \\
        & Vowpal Wabbit \cite{Langford2007VW}      & - & - & \CIRCLE  & - & - & - &
\CIRCLE & - & - \\
        & TensorFlow \cite{tensorflow2015-whitepaper}        & \CIRCLE & -  & \CIRCLE  & \LEFTcircle & - &
\LEFTcircle & - & \LEFTcircle & - \\
        & PyTorch \cite{NEURIPS2019_9015} & \CIRCLE & - & \CIRCLE & \LEFTcircle
& \LEFTcircle & - & - & \LEFTcircle & - \\
        & Caffe \cite{jia2014caffe}           & \CIRCLE & -  & \CIRCLE & \LEFTcircle & \LEFTcircle
& - & - & \LEFTcircle & \CIRCLE \\
        & Keras \cite{chollet2015keras}            & \CIRCLE & -  & \CIRCLE & \LEFTcircle & \LEFTcircle
& - & - & \LEFTcircle & \CIRCLE \\
        & scikit-learn \cite{pedregosa2011scikit}       & \LEFTcircle & - & \LEFTcircle  & \LEFTcircle & -
& - & - & \LEFTcircle & - \\
        & SciPy \cite{jones2014scipy}             & \CIRCLE & \CIRCLE  & -  &
\CIRCLE & - & - & - & \LEFTcircle & \CIRCLE \\
        & MATLAB \cite{matlab_fminsearch}            & \CIRCLE & \CIRCLE & - &
\CIRCLE & - & - & - & \LEFTcircle & - \\
        & Julia (\texttt{Optim.jl}) \cite{mogensen2018optim}         &
\CIRCLE & - & - & \CIRCLE & - & - & - & \CIRCLE & - \\
        \cmidrule[1pt]{2-11}
    \end{tabular}
\caption{
Feature comparison: \CIRCLE~= provides feature,
\LEFTcircle~= partially provides feature, -~= does not provide feature.
{\it unified framework} indicates if there is a form of generic/unified
optimization framework; {\it constraints} and {\it separable functions /
batches} indicate support for constrained functions and separable functions;
{\it arbitrary functions} means arbitrary objective functions are easily
implemented; {\it arbitrary optimizers} means arbitrary optimizers are easily
implemented; {\it sparse gradient} indicates that the framework can natively
take advantage of sparse gradients; {\it categorical} refers to if support
for categorical features exists; {\it arbitrary types} mean that arbitrary types
can be used for the parameters $x$;
{\it callbacks} indicates that user-implementable callback support is available.
}
\label{tab:comparison}
\vspace{1ex}
\hrule
\end{table}

\section{Overview}
\label{sec:overview}



%

The task of optimizing an objective function with {\tt ensmallen} is
straightforward.  The class of objective function (e.g., arbitrary, constrained,
differentiable, etc.) defines the implementation requirements.
Each objective function type has a minimal set of methods that must be implemented.
Typically this is only between one and four methods.
As an example,
to optimize an objective function $f(x)$ that is differentiable,
implementations of $f(x)$ and $f'(x)$ are required.
One of the optimizers for differentiable functions,
such as L-BFGS~\cite{liu1989limited},
can then be immediately employed.

Whenever possible, {\tt ensmallen} will automatically infer methods that are
not provided.  For instance, given a separable objective function $f(x) = \sum_i
f_i(x)$ where an implementation of $f_i(x)$ is provided (as well as the number
of such separable objectives), an implementation of $f(x)$ can be automatically
inferred.  This is done at compile-time, and so there is no additional runtime
overhead compared to a manual implementation.  C++ template metaprogramming
techniques~\cite{abrahams2004c++,alexandrescu2001modern, veldhuizen1998c++} are
internally used to produce efficient code during compilation.

Not every type of objective function can be used with every type of optimizer.
For instance, since L-BFGS is a differentiable optimizer,
it cannot be used with a non-differentiable object function type
(e.g. an arbitrary function).
When an optimizer is used with a user-provided objective function,
an internal mechanism automatically checks the requirements,
resulting in user-friendly error messages if any required methods are not detected.

\subsection{Types of Objective Functions}

In most cases, the objective function $f(x)$ has inherent attributes;
for example, as mentioned in the previous paragraphs, $f(x)$ might be differentiable.
The internal framework in {\tt ensmallen} can optionally take advantage of such attributes.
In the example of a differentiable function $f(x)$,
the user can provide an implementation of the gradient $f'(x)$,
which in turn allows a first-order optimizer to be used.  This generally leads
to significant speedups when compared to using only $f(x)$.
To allow exploitation of such attributes, the optimizers are built to work with
many types of objective functions.  The classes of objective functions
are listed below.

For details on the required signatures for each objective function type
(such as {\tt DifferentiableFunctionType}),
see the online documentation at \mbox{\url{https://ensmallen.org/docs.html}}.

\begin{itemize}
\item {\bf Arbitrary functions} ({\tt ArbitraryFunctionType}).
No assumptions are made on function $f(x)$ and only the objective
$f(x)$ can be computed for a given $x$.

\item {\bf Differentiable functions} ({\tt DifferentiableFunctionType}).
A differentiable function $f(x)$ is an arbitrary function whose gradient $f'(x)$
can be computed for a given $x$, in addition to the objective.

\item {\bf Partially differentiable functions} ({\tt
PartiallyDifferentiableFunctionType}).  A partially differentiable function
$f(x)$ is a differentiable function with the additional property that the
gradient $f'(x)$ can be decomposed along some basis $j$ such that $f_j'(x)$ is
sparse.  Often, this is used for coordinate descent algorithms (i.e., $f'(x)$
can be decomposed into $f_{1}'(x)$, $f_{2}'(x)$, etc.).

\item {\bf Arbitrary separable functions} ({\tt
ArbitrarySeparableFunctionType}).  An arbitrary separable function is an
arbitrary function $f(x)$ that can be decomposed into the sum of several
objective functions: $f(x) = \sum\nolimits_i f_i(x)$.

\item {\bf Differentiable separable functions} ({\tt
DifferentiableSeparableFunctionType}).  A differentiable separable function is a
separable arbitrary function $f(x)$ where the individual gradients $f_i'(x)$ are
also computable.

\item {\bf Categorical functions} ({\tt CategoricalFunctionType}).  A
categorical function type is an arbitrary function $f(x)$ where some (or all)
dimensions of $x$ take discrete values from a set.

\item {\bf Constrained functions} ({\tt ConstrainedFunctionType}).
A constrained function $f(x)$ is a differentiable function%
\footnote{Generally, constrained functions  do not need to be differentiable.
However, this is a requirement here, as all of the current optimizers in {\tt ensmallen}
for constrained functions require a gradient to be available.}
subject to constraints of the form $c_i(x)$; when the constraints are satisfied, $c_i(x) = 0\; \forall \; i$.
Minimizing $f(x)$ then means minimizing $f(x) + \sum_i c_i(x)$.

\item {\bf Semidefinite programs} (SDPs).  {\it (These are a subset of
constrained functions.)}  {\tt ensmallen} has special
support to make optimizing semidefinite
programs~\cite{vandenberghe1996semidefinite} straightforward.
\end{itemize}

\subsection{Pre-Built Optimizers}

For each class of the objective functions,
{\tt ensmallen} provides a set of pre-built optimizers:

\begin{itemize}
  \item {\bf For arbitrary functions:}  Simulated
annealing~\cite{kirkpatrick1983optimization}, CNE
(Conventional Neural Evolution)~\cite{montana1989training}, DE (Differential
Evolution)~\cite{storn1997differential}, PSO (Particle Swarm
Optimization)~\cite{Kennedy1995}, SPSA (Simultaneous Perturbation
Stochastic Approximation)~\cite{spall1992multivariate}.

  \item {\bf For differentiable functions:}  L-BFGS~\cite{liu1989limited},
Frank-Wolfe~\cite{jaggi2013revisiting}, gradient descent.

  \item {\bf For partially differentiable functions.}  SCD (Stochastic
Coordinate Descent)~\cite{Shalev-Shwartz2009}.

  \item {\bf For arbitrary separable functions:}  CMA-ES (Covariance Matrix
Adaptation Evolution Strategy)~\cite{Hansen2001}.

  \item {\bf For differentiable separable functions:}
AdaBound~\cite{Luo2019AdaBound},
AdaDelta~\cite{zeiler2012adadelta}, AdaGrad~\cite{duchi2011adaptive},
Adam~\cite{Kingma2014}, AdaMax~\cite{Kingma2014},
AMSBound~\cite{Luo2019AdaBound}, AMSGrad~\cite{reddi2019convergence},
Big Batch SGD~\cite{De2017}, Eve~\cite{Koushik2016}, FTML (Follow The Moving
Leader)~\cite{Zheng2017},
Hogwild!~\cite{recht2011hogwild}, IQN
(Incremental Quasi-Newton)~\cite{1106.5730}, Katyusha~\cite{Allen-Zhu2016},
Lookahead~\cite{Zhang2019}, SGD with momentum~\cite{rumelhart1988learning},
Nadam~\cite{Dozat2015},
NadaMax~\cite{Dozat2015}, SGD with Nesterov momentum~\cite{Nesterov1983},
Optimistic
Adam~\cite{daskalakis2017training}, QHAdam (Quasi-Hyperbolic
Adam)~\cite{ma2019qh}, QHSGD
(Quasi-Hyperbolic Stochastic Gradient Descent)~\cite{ma2019qh},
RMSProp~\cite{tieleman2012lecture},
SARAH/SARAH+~\cite{Nguyen2017}, stochastic gradient descent, SGDR (Stochastic Gradient
Descent with Restarts)~\cite{Loshchilov2016}, Snapshot SGDR~\cite{Huang2017},
SMORMS3~\cite{Funk2015}, SVRG (Stochastic Variance Reduced
Gradient)~\cite{Johnson2013}, SWATS~\cite{Keskar2017},
SPALeRA (Safe Parameter-wise Agnostic LEarning Rate
Adaptation)~\cite{Schoenauer2017},
WNGrad~\cite{Wu2018}.

  \item {\bf For categorical functions:}  Grid search.

  \item {\bf For constrained functions:}  Augmented Lagrangian method,
primal-dual interior point SDP solver, LRSDP (low-rank accelerated SDP
solver)~\cite{burer2003nonlinear}.
\end{itemize}

\subsection{Example Usage}
\label{sec:linreg_example}

Let us consider the problem of linear regression, where we are
given a matrix of predictors $\bm X \in \mathcal{R}^{n \times d}$ and a vector
of responses $\bm y \in \mathcal{R}^n$.  Our task is to find the best linear
model $\bm \theta \in \mathcal{R}^d$; that is, we want to find
$\bm \theta^* = \operatornamewithlimits{argmin}_{\bm\theta} f(\bm \theta)$ for
\begin{equation}
f(\bm \theta) = \| \bm X \bm \theta - \bm y \|^2 = (\bm X \bm \theta - \bm y)^T
(\bm X \bm \theta - \bm y).
\label{eqn:obj_lr}
\end{equation}

\noindent
From this we can derive the gradient $f'(\bm \theta)$:
\begin{equation}
f'(\bm \theta) = 2 \bm X^T (\bm X \bm \theta - \bm y).
\label{eqn:grad_lr}
\end{equation}

To find $\bm \theta^*$ using a differentiable
optimizer\footnote{Typically, in practice, solving a linear regression model can
be done directly by taking the pseudoinverse: $\theta^* = X^\dagger y$.  However,
the objective function is easy to describe and useful for demonstration, so we
use it in this document.}, we simply need to provide implementations of $f(\bm \theta)$ and
$f'(\bm \theta)$ according to the signatures required by the {\tt
DifferentiableFunctionType} of objective function.  For a differentiable
function, only two methods are necessary: {\tt Evaluate()} and {\tt Gradient()}.
The pre-built L-BFGS optimizer can be used to find $\bm \theta^*$.

Figure~\ref{fig:lr_function} shows an example implementation.
We hold {\tt X} and {\tt y} as members of the
{\tt LinearRegressionFunction class},
and {\tt theta} is used to represent $\bm \theta$.
Via the use of Armadillo~\cite{sanderson2016armadillo},
the linear algebra expressions to implement the objective function and gradient
are readable in a way that closely matches Equations~(\ref{eqn:obj_lr}) and~(\ref{eqn:grad_lr}).


\begin{figure}[t!]
\hrule
\vspace{1ex}
\centering
\begin{minted}[fontsize=\small]{c++}
#include <ensmallen.hpp>

class LinearRegressionFunction
{
 public:
  LinearRegressionFunction(const arma::mat& in_X, const arma::vec& in_y) : X(in_X), y(in_y)  { }

  double Evaluate(const arma::mat& theta)  { return (X * theta - y).t() * (X * theta - y); }

  void Gradient(const arma::mat& theta, arma::mat& gradient)  { gradient = 2 * X.t() * (X * theta - y); }

 private:
  const arma::mat& X;
  const arma::vec& y;
};

int main()
{
  arma::mat X;
  arma::vec y;
  
  // ... set the contents of X and y here ...
  
  ens::LinearRegressionFunction f(X, y);

  ens::L_BFGS optimizer; // create the optimizer with default parameters

  arma::mat theta_best(X.n_rows, 1, arma::fill::randu);  // initial starting point (uniform random values)

  optimizer.Optimize(f, theta_best);
  // at this point theta_best contains the best parameters
}
\end{minted}
\hrule
\vspace*{-0.5em}
\caption{An example implementation of an objective function class for linear
regression and usage of the L-BFGS optimizer in {\tt ensmallen}.
The online documentation for all ensmallen optimizers
is at \mbox{\url{https://ensmallen.org/docs.html}}.
The {\tt arma::mat} and {arma::vec} types are 
dense matrix and vector classes
from the Armadillo linear algebra library~\cite{sanderson2016armadillo},
with the corresponding online documentation at \mbox{\url{http://arma.sf.net/docs.html}}.
}
\label{fig:lr_function}
\end{figure}

\section{Advanced Usage}
\label{sec:advanced}

In addition to the support for various objective functions and
the provision of many pre-built optimizers covered in Section~\ref{sec:overview},
the {\tt ensmallen} library provides further functionality, suitable for advanced users.
This includes:

\begin{itemize}

\item
Inference of functions not supplied by the user.
For example, a user might supply a single function named
{\tt EvaluateWithGradient()}
in order to allow sharing common computation that may be 
present in {\tt Evaluate()} and {\tt Gradient()}.
Those two functions can then be automatically inferred when needed.

\item
Callbacks, which can specify custom behavior during the optimization.
Examples include printing the loss function value at each iteration,
terminating when a time limit is reached,
changing the step size,
and adding custom constraints when they are violated by the current solution.

\item
Use of types other than {\tt arma::mat} when calling {\tt Optimize()}.
This can include
integer-valued matrices ({\tt arma::imat}),
sparse matrices ({\tt arma::sp\_mat}),
or any type whose implementation matches the Armadillo API~\cite{sanderson2016armadillo,sanderson2018user}.

\end{itemize}

Each of the above points is covered in more detail in the following subsections.

  \subsection{Automatically Inferring Missing Methods}
\label{sec:automatic}

In Section~\ref{sec:linreg_example}, we saw an example of how a user might implement {\tt
Evaluate()} and {\tt Gradient()} for the linear regression objective function
and use {\tt ensmallen} to find the minimum.
However, there is an inefficiency:
the objective function computation is defined as $f(\bm \theta) = \| \bm X \bm \theta - \bm y \|^2$,
and the gradient computation is defined as $f'(\bm \theta) = 2 \bm X^T (\bm X \bm \theta - \bm y)$.
There is a shared inner computation in $f(\bm \theta)$ and $f'(\bm \theta)$: the
term $(\bm X \bm \theta - \bm y)$.
If $f(\bm \theta)$ and $f'(\bm \theta)$ are implemented as separate functions,
there is no easy way to exploit this shared computation.

The differentiable optimizers in {\tt ensmallen} treat the given functions as oracular,
and do not know anything about the internal computations of the functions.
This inefficiency%
\footnote
  {It may be possible to use a specialized implementation of
  auto-differentiation or a programming language with introspection that
  operates directly on the abstract syntax tree~\cite{Jones_AST_2003} of the
  given objective and gradient computations to avoid this inefficiency and
  successfully share the computations across the gradient and objective.
  However, at the time of this writing, we are not aware of any optimization
  packages that actually do this.
  }
can apply to any optimization package that accepts an objective
function and its gradient as separate parameters,
such as SciPy, {\tt Optim.jl} and {\tt bfgsmin()} from Octave~\cite{octave}.

To work around this issue, {\tt ensmallen} internally uses template metaprogramming techniques to allow
the user to provide {\it either} separate implementations of the objective
function and gradient, {\it or} a combined implementation that computes {\it
both} the objective function and gradient simultaneously.
The latter options allows the sharing of inner computations.
That is, the user can provide the methods {\tt Evaluate()} and {\tt Gradient()},
or {\tt EvaluateWithGradient()}.
For the example objective function above,
we empirically show (Section~\ref{sec:experiments}) that the ability to provide
{\tt EvaluateWithGradient()} can result in a significant speedup.
Similarly, when implementing a differentiable optimizer in {\tt ensmallen},
it is possible to use {\it either} {\tt Evaluate()} and {\tt Gradient()},
{\it or} {\tt EvaluateWithGradient()} during optimization.

The same technique is used to infer and provide more missing methods than
just {\tt EvaluateWithGradient()} or {\tt Evaluate()} and {\tt Gradient()}.  
For instance, separable functions, differentiable separable functions, constrained
functions, and categorical functions each have inferrable methods.  Not all of
these possibilities are currently implemented in {\tt ensmallen}, but the
existing framework makes it straightforward to add more.  Below are
examples that are currently implemented:

\begin{itemize}
  \item {\it (Differentiable functions.)}  If the user provides an objective
function with {\tt Evaluate()} and {\tt Gradient()}, we can automatically
synthesize {\tt EvaluateWithGradient()}.

  \item {\it (Differentiable functions.)}  If the user provides an objective
function with {\tt EvaluateWithGradient()}, we can synthesize {\tt Evaluate()}
and/or {\tt Gradient()}.

  \item {\it (Separable functions.)}  If the user provides an objective
function with {\tt Evaluate()} and {\tt NumFunctions()}, we can sythesize a
non-separable version of {\tt Evaluate()}.

  \item {\it (Separable functions.)}  If the user provides an objective function
with {\tt Gradient()} and {\tt NumFunctions()}, we can synthesize a
non-separable version of {\tt Gradient()}.

  \item {\it (Separable functions.)}  If the user provides an objective
function with {\tt EvaluateWithGradient()} and {\tt NumFunctions()}, we can
synthesize a non-separable version of {\tt Gradient()}.
\end{itemize}

For more precise details on exactly how the method generation works,
see Appendix~\ref{sec:automatic_details},
where we describe the framework in a simplified form,
focusing only the {\tt EvaluateWithGradient()}/{\tt Evaluate()}/{\tt Gradient()}
example described above.

  \subsection{Callbacks}
\label{sec:callbacks}

Many of the optimizers in {\tt ensmallen} offer the ability to monitor
and modify parts of the optimization process.
Example modifications include changing the step size,
adding custom constraints when they are violated by the current solution,
or providing custom heuristics to find and investigate feasible solutions.

In many existing toolkits, this type of functionality is provided only via
solver-specific interfaces.  For instance, the {\tt tick} statistical learning
toolkit~\cite{bacry2017tick} requires the use of a solver-specific {\tt History}.
In contrast, {\tt ensmallen} provides optimizer-independent callbacks to allow
classes to inspect and work with internal parts of the optimization process.
In particular, the callbacks allow code to be
executed regularly during an optimization session.

\begin{table}[t!]
\centering
\small
\begin{tabular}{lll}
\toprule
{\bf Function} & {\bf Description} & {\bf Function type} \\
\hline
\texttt{BeginOptimization}   & Called at the beginning of the optimization process  & {\it all} \\
\texttt{EndOptimization}     & Called at the end of the optimization process & {\it all} \\
\texttt{Evaluate}            & Called after any call to {\tt Evaluate()}            & Arbitrary, Differentiable, Partially differentiable,  \\
                             &                                                      & Arbitrary separable, Differentiable separable \\
\texttt{EvaluateConstraint}  & Called after any call to {\tt EvaluateConstraint()}  & Constrained \\
\texttt{Gradient}            & Called after any call to {\tt Gradient()}            & Differentiable, Partially differentiable \\
                             &                                                      & Differentiable separable \\
\texttt{GradientConstraint}  & Called after any call to {\tt GradientConstraint()}  & Constrained \\
\texttt{BeginEpoch}          & Called at the beginning of a pass over the data      & Differentiable separable \\
\texttt{EndEpoch}            & Called at the end of a pass over the data            & Differentiable separable \\

\bottomrule
\end{tabular}
\vspace{0.5ex}
\caption
  {
  Available callback routines, with brief descriptions.
  Optional additional arguments have been omitted for brevity.
  See {\href{http://www.ensmallen.org/docs.html}{\mbox{\tt http://www.ensmallen.org/docs.html}}} for more detailed documentation.
  }
\label{tab:callback_list}
\end{table}

\begin{figure}[b!]
\centering
\hrule
\vspace{1ex}
\begin{minted}[fontsize=\small]{c++}
RosenbrockFunction f;
arma::mat coordinates = f.GetInitialPoint();

MomentumSGD optimizer(0.01, 32, 100000, 1e-5, true, MomentumUpdate(0.5));
optimizer.Optimize(f, coordinates, EarlyStopAtMinLoss(), ProgressBar());
\end{minted}
\hrule
\vspace*{-0.5em}
\caption
  {
  Code snippet to demonstrate usage of the callback functionality using
pre-defined callbacks: \texttt{EarlyStopAtMinLoss} and \texttt{ProgressBar}.
{\tt EarlyStopAtMinLoss} will terminate the optimization as soon as the
objective value stars increasing, and {\tt ProgressBar} will print a progress
bar during the optimization.
  }
\label{fig:example_prog_callbacks}
\end{figure}

\subsubsection{Using Callbacks}

To use callbacks, either for optimization, tuning or logging, 
arbitrary callbacks to any optimizer can be optionally provided
to the {\tt Optimize()} function.
Figure~\ref{fig:example_prog_callbacks} contains a code snippet which
briefly demonstrates usage of the callback functionality.  Given the pre-defined
callbacks {\tt EarlyStopAtMinLoss} and {\tt ProgressBar}, the code snippet shows
not only how the {\tt MomentumSGD} optimizer can be used to find the best
coordinates but also how the callbacks can be used to control and monitor the
optimization.

Eight types of optimization callback routines are available,
as shown in Table~\ref{tab:callback_list}.
These callbacks are regularly called during the optimization process,
depending on the objective function type.
Callbacks are executed in the order that they are specified.
All callbacks except for {\tt BeginOptimization()} and {\tt EndOptimization()}
may terminate the optimization via their {\tt bool} return values,
where {\tt true} indicates that the optimization should be stopped.
By default, subsequent callbacks are not called if an earlier callback terminates
the optimization: the optimizer terminates immediately.

There are several pre-built callbacks that can be used without
needing any custom code:

\begin{itemize}
  \item {\tt EarlyStopAtMinLoss}: stops the optimization process if the loss
stops decreasing or no improvement has been made.

  \item {\tt PrintLoss}: callback that prints loss to {\tt stdout} or a
specified output stream.

  \item {\tt ProgressBar}: callback that prints a progress bar to {\tt stdout}
or a specified output stream.

  \item {\tt StoreBestCoordinates}: callback that stores the model parameter
after every epoch if the objective decreased.
\end{itemize}

Note that to use the {\tt StoreBestCoordinates} callback, the user will need to
instantiate a {\tt StoreBestCoordinates} object, and then call {\tt
BestCoordinates()} in order to recover the best coordinates found during the
optimization.  This process is detailed more in the following section.

\begin{figure}[t!]
\centering
\hrule
\vspace{1ex}
\begin{minted}[fontsize=\small]{c++}
class CustomCallback
{
 public:
  template<typename OptimizerType, typename FunctionType, typename MatType>
  bool Evaluate(OptimizerType& opt, FunctionType& function, const MatType& coordinates, const double objective)
  {
    std::cout << "The optimization process called Evaluate()!" << std::endl;
    return false; // Do not terminate, continue the optimization process.
  }
};
\end{minted}
\hrule
\vspace*{-0.5em}
\caption
  {
  An example of a custom callback.  This callback prints to {\tt std::cout}
after each time the {\tt Evaluate()} function is called by the optimizer. In the example the
callback always returns {\tt false}, meaning that the optimization should not be
terminated on behalf of the callback.
  }
\label{fig:example_prog_callbacks_2}
\end{figure}

\begin{figure}[t!]
\centering
\hrule
\vspace{1ex}
\begin{minted}[fontsize=\small]{c++}
struct CustomCallback
{
  CustomCallback(double rIn) : r(rIn) {}

  template<typename OptimizerType, typename FunctionType, typename MatType>
  bool StepTaken(OptimizerType& optimizer, FunctionType& function, MatType& coordinates)
  {
    // Multiply the step size by r (hopefully r is less than 1!).
    optimizer.StepSize() *= r;
    return false; // Do not terminate the optimization.
  }

  double r;
};

RosenbrockFunction f;
arma::mat coordinates = f.GetInitialPoint();

Adam opt;
CustomCallback cb(0.9); // Instantiate the custom callback...
opt.Optimize(f, coordinates, cb); // ...and call Optimize() with that object!
\end{minted}
\hrule
\vspace*{-0.5em}
\caption
  {
  Code snippet demonstrating how to add additional parameters/state to a
callback and accessing optimizer-specific parameters.
  }
\label{fig:example_prog_callbacks_parameter}
\end{figure}

\subsubsection{Custom Callbacks}

Implementing a custom callback is straightforward:
the only requirement is a class that has functions whose names are the same as the callback
functions listed in Table~\ref{tab:callback_list}.
Comprehensive documentation on the required signatures for each callback can be
found in the {\tt ensmallen} documentation at
\url{http://ensmallen.org/docs.html#callback-states}.

If a custom callback is desired to take an action before the optimization process starts, then only a
{\tt CustomCallback::BeginOptimization()} method is required to perform the desired action.
An example of a custom callback that 
prints a line to {\tt std::cout} every time the optimization calls
{\tt Evaluate()} is shown in Figure~\ref{fig:example_prog_callbacks_2}.
The {\tt CustomCallback} class can be used exactly like {\tt EarlyStopAtMinLoss}
in Figure~\ref{fig:example_prog_callbacks}---the only change
needed is to add {\tt CustomCallback()} to the list of arguments passed to
{\tt optimizer.Optimize()}.

If a callback class requires additional parameters or state beyond what is
passed through the predefined arguments to functions in
Table~\ref{tab:callback_list}, a user should manually create an instance of the
callback class with those additional parameters, and then pass the object to the
optimizer as a callback when {\tt Optimize()} is called.  {\tt ensmallen} does
not modify or dereference the object, so it is safe to use for this purpose.
Figure~\ref{fig:example_prog_callbacks_parameter} provides an example,
where we pass an instantiated custom callback that takes an additional
step-size decay parameter as input.
In addition, inside the {\tt StepTaken()} callback, we use the
optimizer's interface function ({\tt StepSize()}) to update the step size.
Note that  if this callback is attempted to be used with an optimizer that
did not have a {\tt StepSize()} function, compilation would fail.
As such, some care is necessary when implementing custom callbacks.

There is no performance penalty if no callbacks are used.
Figure~\ref{fig:callback_compilter_opt} shows two programs;
one explicitly without callbacks, and one with an empty callback.
The implementation of callback facility is internally
done via template metaprogramming.
In all cases, modern C++ compilers (e.g. {\tt clang-1100.0.33.16} and {\tt g++} 9.2.1)
optimized away the unused code;
the resultant machine code appears as if the callback code never existed
in the first place.
Both programs produce the exact same machine code,
indicating that there is no performance penalty if no callbacks are used.

\begin{figure}[t!]
\centering
\hrule
\vspace{1ex}
\begin{minipage}{0.47\textwidth}
\begin{minted}[fontsize=\small,stripnl=false]{c++}
struct Optimizer
{
  template<typename FT>
  void Optimize(FT& f, arma::mat& p)
  {
    f.Evaluate(p);
  }
};


int main()
{
  RosenbrockFunction rf;
  arma::mat parameters = rf.GetInitialPoint();
  Optimizer opt;
  opt.Optimize(rf, parameters);
  return 0;
}

\end{minted}
\end{minipage}
\hfill
\vline
\hfill
\begin{minipage}{0.51\textwidth}
\begin{minted}[fontsize=\small,escapeinside=||]{c++}
struct Optimizer
{
  template<typename FT, |\colorbox{yellow}{typename... CallbackType}|>
  void Optimize(FT& f, arma::mat& p, CallbackType&&... c)
  {
    |\!\!\colorbox{yellow}{Callback::BeginOptimization(*this, f, p, c...);}|
    f.Evaluate(iterate);
  }
};

int main()
{
  RosenbrockFunction rf;
  arma::mat parameters = rf.GetInitialPoint();
  Optimizer opt;
  opt.Optimize(rf, parameters);
  return 0;
}
\end{minted}
\end{minipage}
\vspace{1ex}
\hrule
\caption
  {
  Left panel: A C++ program that mimics the ensmallen optimizer interface
  without any callback functionality. Right panel: A corresponding C++ program
  using an empty callback routine that is automatically optimized out. Both
  programs produce the exact same machine code, resulting in no performance
  penalty if no callbacks are used.
  }
\label{fig:callback_compilter_opt}
\end{figure}

Details on the internal implementation of the callback system can be found in
Appendix~\ref{sec:callback_details}.

  \subsection{Optimization With Various Matrix Types}
\label{sec:templated_optimize}


\begin{figure}[b!]
\hrule
\vspace{1ex}
\begin{minted}[fontsize=\small]{c++}
#include <ensmallen.hpp>

// A trivial parabolic function.  The minimum is at the origin.  The function
// works with any matrix type.
class ParabolicFunction
{
 public:
  template<typename MatType>
  typename MatType::elem_type Evaluate(const MatType& coordinates)
  {
    // Return the sum of squared coordinates.
    return arma::accu(arma::square(coordinates));
  }
};

int main()
{
  // Use simulated annealing to optimize the ParabolicFunction.
  ParabolicFunction pf;
  ens::SA<> optimizer(ens::ExponentialSchedule());

  // element type is double
  arma::mat doubleCoordinates(10, 1, arma::fill::randu);
  optimizer.Optimize(pf, doubleCoordinates);

  // element type is float
  arma::fmat floatCoordinates(10, 1, arma::fill::randu);
  optimizer.Optimize(pf, floatCoordinates);

  // element type is int
  arma::imat intCoordinates(10, 1, arma::fill::randi);
  optimizer.Optimize(pf, intCoordinates);
}
\end{minted}
\hrule
\vspace*{-0.5em}
\caption{Example {\tt ensmallen} program showing that {\tt ensmallen}'s
optimizers can be used to seamlessly optimize functions with many different
element types.}
\label{fig:many_optimize}
\end{figure}

\begin{figure}[t!]
\hrule
\vspace{1ex}
\begin{minted}[fontsize=\small]{c++}
class GradientDescent
{
  template<typename FunctionType, typename MatType, typename GradType = MatType>
  typename MatType::elem_type Optimize(FunctionType& function, MatType& coordinates)
  {
    // The step size is hardcoded to 0.01, and the number of iterations is 1000.
    for (size_t i = 0; i < 1000; ++i)
    {
      GradType gradient;
      function.Gradient(coordinates, gradient);

      // Take the step.
      coordinates -= 0.01 * gradient;
    }

    // Compute and return the final objective.
    return function.Evaluate(coordinates);
  }
};
\end{minted}
\hrule
\vspace*{-0.5em}
\caption{Example implementation of a simple gradient descent optimizer.
For the sake of brevity, functionality such as the ability to configure the parameters has
been deliberately omitted.
The actual {\tt GradientDescent} optimizer in {\tt ensmallen} provides more functionality.
}

\label{fig:gd}
\end{figure}

In the example shown in Section~\ref{sec:linreg_example},
where we introduced the class {\tt LinearRegressionFunction},
the matrix and vector objects are hardcoded as {\tt arma::mat} and {\tt arma::vec}.
These objects hold elements with the C++ type {\tt double},
representing double-precision floating point~\cite{Goldberg_CSUR_1991}.
However, in many applications it can be very important to specify a different
underlying element type (e.g. the single-precision {\tt float}, used by {\tt arma::fmat} and {\tt arma::fvec}).
For instance, in the field of machine learning, neural
networks have been shown to be effective with low-precision floating point
representations for weights~\cite{vanhoucke2011improving}.
Furthermore, many optimization problems have parameters
that are best represented as sparse data~\cite{van2011sparse, recht2011hogwild},
which is represented in Armadillo as the {\tt sp\_mat} class~\cite{sanderson2018user, mca24030070}.
Even alternate representations such as data held on the GPU can be quite
important: the use of GPUs can often result in significant
speedups~\cite{oh2004gpu, athanasopoulos2011gpu}.

In order to handle this diverse set of needs, {\tt ensmallen} has been built in
such a way that any underlying storage type can be used to represent the
coordinates to be optimized---so long as it matches the Armadillo API.
This means that the Bandicoot GPU matrix library\footnote{\url{https://gitlab.com/conradsnicta/bandicoot-code}}
can be used as a drop-in replacement once it is stable.

An example of seamlessly using {\tt ensmallen}'s optimizers with different
underlying storage types is given in Figure~\ref{fig:many_optimize}.  In this
example, the types {\tt arma::mat} (which holds {\tt double}), {\tt arma::fmat}
(which holds {\tt float}), and {\tt arma::imat} (which holds {\tt int}) are all
used with {\tt ensmallen}'s simulated annealing implementation to optimize a
very simple parabolic function.  Due to the use of templates both inside {\tt
ensmallen} and in the implementation of the {\tt ParabolicFunction} class, it is
trivial to change the underlying type used for storage.

Consider the simplified gradient descent optimizer shown in Figure~\ref{fig:gd}.
The use of the template types {\tt FunctionType}, {\tt MatType}, and {\tt
GradType} means that at compilation time, the correct types are substituted in
for {\tt FunctionType}, {\tt MatType}, and {\tt GradType} (which by default is
set to be the same as {\tt MatType} in this code).  So long as each type has all
the methods that are used inside of {\tt Optimize()}, there will be no
compilation problems.  Templates are a technique for code generation; in this
case, that means the code generated will be exactly the same as if {\tt
Optimize()} was written with the types specified in those template parameters.
This means that there is no additional runtime overhead when a different {\tt
MatType} is used.

Appendix~\ref{sec:templated_optimize_details} contains details on the internal
compile-time checks used for providing users with concise error messages, 
avoiding the long errors typically associated with template metaprogramming.

\section{Experiments}
\label{sec:experiments}

To show the efficiency of mathematical optimization with {\tt ensmallen}, we
compare its performance with several other commonly used optimization
frameworks, including some that use automatic differentiation.

\subsection{Simple Optimizations and Overhead}

For our first experiment, we aim to capture the overhead involved in various
optimization toolkits.  In order to do this, we consider the simple and popular
Rosenbrock function~\cite{Rosenbrock1960}:
\begin{equation}
f([x_1, x_2]) = 100 (x_2 - x_1^2)^2 + (1 - x_1^2).
\end{equation}

This objective function is useful for this task because the computational effort
involved in computing $f(\cdot)$ is trivial.  Therefore, if we hold the number
of iterations of each toolkit constant, then this will help us understand the
overhead costs of each toolkit.  For the optimizer, we use simulated
annealing~\cite{kirkpatrick1983optimization}, a gradient-free optimizer.
Simulated annealing will call the objective function numerous times; for each
simulation we limit the optimizer to 100K objective evaluations.

The code used to run this simulation for {\tt ensmallen} (including the
implementation of the Rosenbrock function) is given in
Figure~\ref{fig:rosenbrock_run}.  Note that the {\tt RosenbrockFunction} is
actually implemented in {\tt ensmallen}'s source code, in the directory {\tt
include/ensmallen\_bits/problems/}.


We compare four frameworks for this task:
\begin{itemize}
\itemsep=-1ex
  \item[{\bf (i)}] {\tt ensmallen},
  \item[{\bf (ii)}] {\tt scipy.optimize.anneal} from SciPy 0.14.1~\cite{jones2014scipy},
  \item[{\bf (iii)}] simulated annealing implementation in {\tt Optim.jl} with Julia 1.0.1~\cite{mogensen2018optim},
  \item[{\bf (iv)}] {\tt samin} in the {\tt optim} package for Octave~\cite{octave}.
\end{itemize}

While another option here might be {\tt simulannealbnd()}
in the Global Optimization Toolkit for MATLAB,
no license was available.
We ran our code on a MacBook Pro i7 2018 with 16GB RAM running macOS 10.14 with clang 1000.10.44.2, Julia version 1.0.1, Python 2.7.15, and Octave 4.4.1.

Our initial implementation for each toolkit, corresponding to the line
``default'' in Table~\ref{tab:rosenbrock_results}, was as simple of an
implementation as possible and included no tuning.  This reflects how a typical
user might interact with a given framework.  Only Julia and {\tt ensmallen} are
compiled, and thus are able to avoid the function pointer dereference for
evaluating the Rosenbrock function and take advantage of inlining and related
optimizations.  The overhead of both {\tt scipy} and {\tt samin} are quite
large---{\tt ensmallen} is nearly three orders of magnitude faster for the same
task.

\begin{figure}[t!]
\hrule
\vspace{1ex}
\begin{minted}[fontsize=\small]{c++}
#include <ensmallen.hpp>

struct RosenbrockFunction
{
  template<typename MatType>
  static typename MatType::elem_type Evaluate(const MatType& x) const
  {
    return 100 * std::pow(x[1] - std::pow(x[0], 2), 2) + std::pow(1 - x[0], 2);
  }
};

int main()
{
  arma::wall_clock clock;

  RosenbrockFunction rf;
  ens::ExponentialSchedule sched;
  // A tolerance of 0.0 means the optimization will run for the maximum number of iterations.
  ens::SA<> s(sched, 100000, 10000, 1000, 100, 0.0);

  // Get the initial point of the optimization.
  arma::mat parameters = rf.GetInitialPoint();

  // Run the optimization and time it.
  clock.tic();
  s.Optimize(rf, parameters);
  const double time = clock.toc();
  std::cout << time << std::endl << "Result (optimal 1, 1): " << parameters.t();
  return 0;
}
\end{minted}
\hrule
\vspace*{-0.5em}
\caption{Code to use {\tt ensmallen} to optimize the Rosenbrock function using
100K iterations of simulated annealing.}
\label{fig:rosenbrock_run}
\end{figure}

\begin{table}[b!]
\begin{center}
\begin{tabular}{lcccc}
\toprule
 & {\tt ensmallen} & {\tt scipy} & {\tt Optim.jl} & {\tt samin} \\
\midrule
default & {\bf 0.004s} & 1.069s & 0.021s & 3.173s \\
tuned & & 0.574s & & 3.122s \\
\bottomrule
\end{tabular}
\end{center}
\vspace*{-0.5em}
\caption{Runtimes for $100$K iterations of simulated annealing with various
frameworks on the simple Rosenbrock function.  Julia code runs do not count
compilation time.  The {\it tuned} row indicates that the code was manually
modified for speed.}
\label{tab:rosenbrock_results}
\end{table}

However, both Python and Octave have routes for acceleration,
such as Numba~\cite{lam2015numba}, MEX bindings and JIT compilation.
We hand-optimized the Rosenbrock implementation using Numba,
which required significant modification of the
underlying \texttt{anneal.anneal()} function.
These techniques did produce some speedup,
as shown in the second row of Table~\ref{tab:rosenbrock_results}.
For Octave, a MEX binding did not produce a noticeable difference.
We were unable to tune either \texttt{ensmallen} or
\texttt{Optim.jl} to get any speedup,
suggesting that novice users will easily be able
to write efficient code in these cases.

\subsection{Large-Scale Linear Regression Problems}

\begin{table}[t!]
\centering
\begin{tabular}{lccccc}
\toprule
{\em algorithm} & $d$: 100, $n$: 1k & $d$: 100, $n$: 10k & $d$: 100, $n$:
100k & $d$: 1k, $n$: 100k \\
\midrule
\texttt{ensmallen-1} & {\bf 0.001s} & {\bf 0.009s} & {\bf 0.154s} & {\bf 2.215s} \\
\texttt{ensmallen-2} & 0.002s & 0.016s & 0.182s & 2.522s \\
\texttt{Optim.jl} & 0.006s & 0.030s & 0.337s & 4.271s \\
\texttt{scipy} & 0.003s & 0.017s & 0.202s & 2.729s \\
\texttt{bfgsmin} & 0.071s & 0.859s & 23.220s & 2859.81s\\
\texttt{ForwardDiff.jl} & 0.497s & 1.159s & 4.996s & 603.106s \\
\texttt{autograd} & 0.007s & 0.026s & 0.210s & 2.673s \\
\bottomrule
\end{tabular}
\vspace*{0.25ex}
\caption{
Runtimes for the linear regression function on various dataset sizes,
with $n$ indicating the number of samples,
and $d$ indicating the dimensionality of each sample.
All Julia runs do not count compilation time.}
\label{tab:lbfgs}
\end{table}

Next, we consider the linear regression example described in
Section~\ref{sec:linreg_example}.  For this task we use the first-order L-BFGS
optimizer~\cite{liu1989limited}, implemented in {\tt ensmallen} as the {\tt
L\_BFGS} class.  Using the same four packages, we implement
the linear regression objective and gradient.  Remembering that {\tt ensmallen}
allows us to share work across the objective function and gradient
implementations (Section~\ref{sec:automatic}), for {\tt ensmallen} we implement
a version with only {\tt EvaluateWithGradient()}, denoted as {\tt ensmallen-1}.
We also implement a version with both \texttt{Evaluate()} and
\texttt{Gradient()}: \texttt{ensmallen-2}.  We also use automatic
differentiation for Julia via the 
\texttt{ForwardDiff.jl}~\cite{RevelsLubinPapamarkou2016} package
and for Python via the \texttt{Autograd}~\cite{maclaurin2015autograd} package.  
For GNU Octave we use the \texttt{bfgsmin()} function.

Results for various data sizes are shown in Table~\ref{tab:lbfgs}.  For each
implementation, L-BFGS was allowed to run for only $10$ iterations and never
converged in fewer iterations.  The datasets used for training are highly noisy random
data with a slight linear pattern. Note that the exact data is not relevant
for the experiments here, only its size.  Runtimes are reported as the
average across 10 runs.

The results indicate that \texttt{ensmallen} with
\texttt{EvaluateWithGradient()} is the fastest approach.
Furthermore, the use of \texttt{EvaluateWithGradient()} yields
non-negligible speedup over the \texttt{ensmallen-2} implementation with
both the objective and gradient implemented separately.  In addition, although
the automatic differentiation support makes it easier for users to write their
code (since they do not have to write an implementation of the gradient), we
observe that the output of automatic differentiators is not as efficient,
especially with \texttt{ForwardDiff.jl}.  We expect this effect to be
more pronounced with increasingly complex objective functions.

\subsection{Easy Pluggability of Various Optimizers}

\begin{figure}[H]
  \vspace*{-1em}
  \centering
  \includegraphics[width=\textwidth,height=0.5\textwidth]{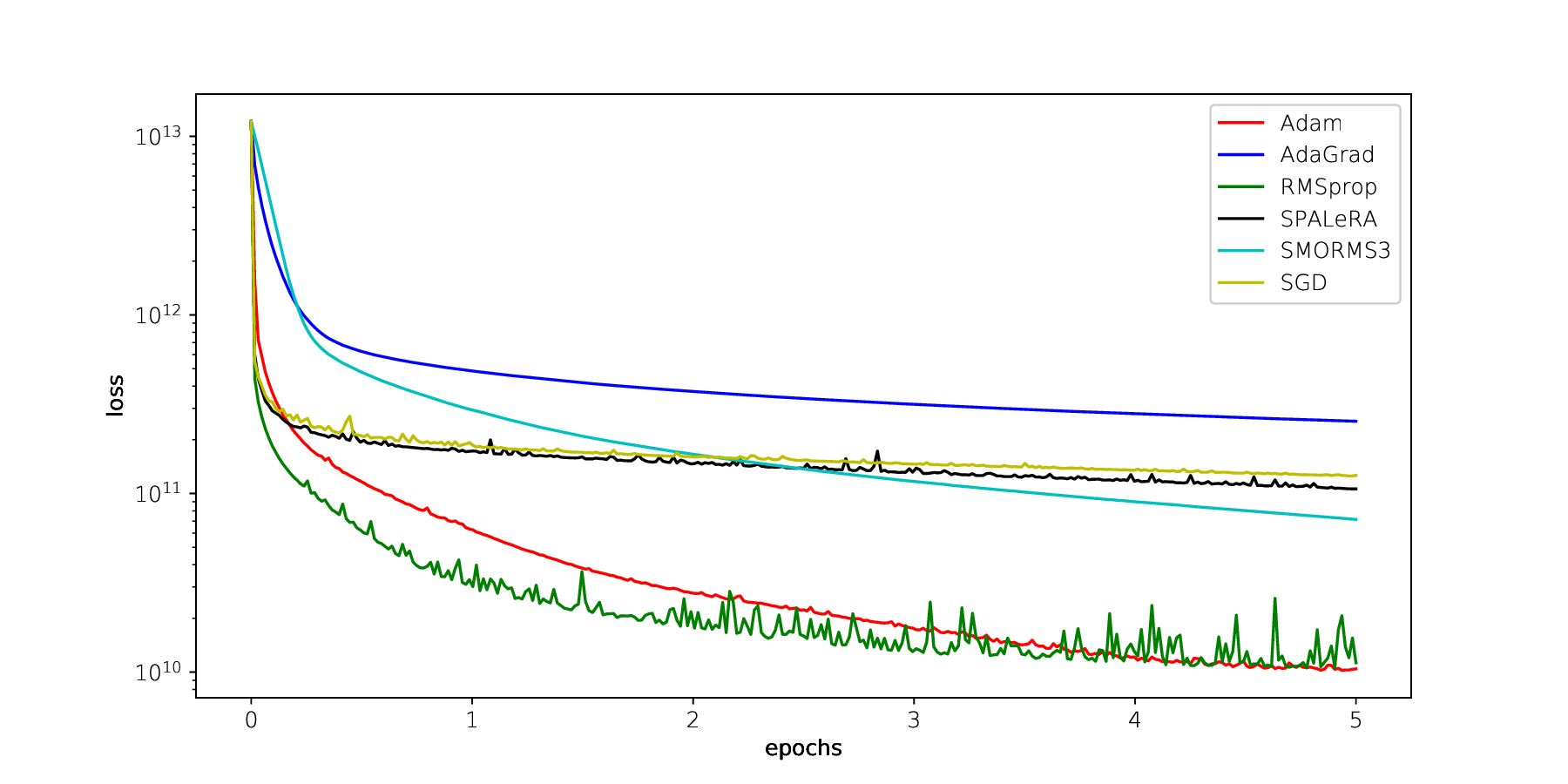}
  \vspace*{-2.5em}
\caption{
Example usage of six {\tt ensmallen} optimizers to optimize a linear regression
function on the {\tt YearPredictionMSD} dataset~\cite{ucimlrepository} for 5
epochs of training.  The optimizers can be tuned for better performance.}
\label{fig:learning_curve}
\end{figure}

Lastly, we demonstrate the easy pluggability in \texttt{ensmallen}
for using various optimizers on the same task.
Using a version of {\tt LinearRegressionFunction} from
Section~\ref{sec:linreg_example} adapted for separable differentiable
optimizers, we run six optimizers with default parameters in just 8 lines of
code, as shown in Fig.~\ref{fig:learning_curve_code}.
Applying these optimizers to the \texttt{YearPredictionMSD}
dataset from the UCI repository~\cite{ucimlrepository}
yields the learning curves shown in Fig.~\ref{fig:learning_curve}.

\begin{figure}[H]
\hrule
\vspace{1ex}
\begin{minted}[fontsize=\small]{c++}
// X and y are data.
LinearRegressionFunction lrf(X, y);

using namespace ens;
StandardSGD<>().Optimize(lrf, sgdModel);
Adam().Optimize(lrf, adamModel);
AdaGrad().Optimize(lrf, adagradModel);
SMORMS3().Optimize(lrf, smorms3Model);
SPALeRASGD().Optimize(lrf, spaleraModel);
RMSProp().Optimize(lrf, rmspropModel);
\end{minted}
\hrule
\vspace*{-0.5em}
\caption{{\tt ensmallen} makes it easy to switch out optimizer types:
8 lines of code run 6 optimizers on one problem.}
\label{fig:learning_curve_code}
\end{figure}

\begin{figure}[H]
  \centering
  \includegraphics[width=\textwidth]{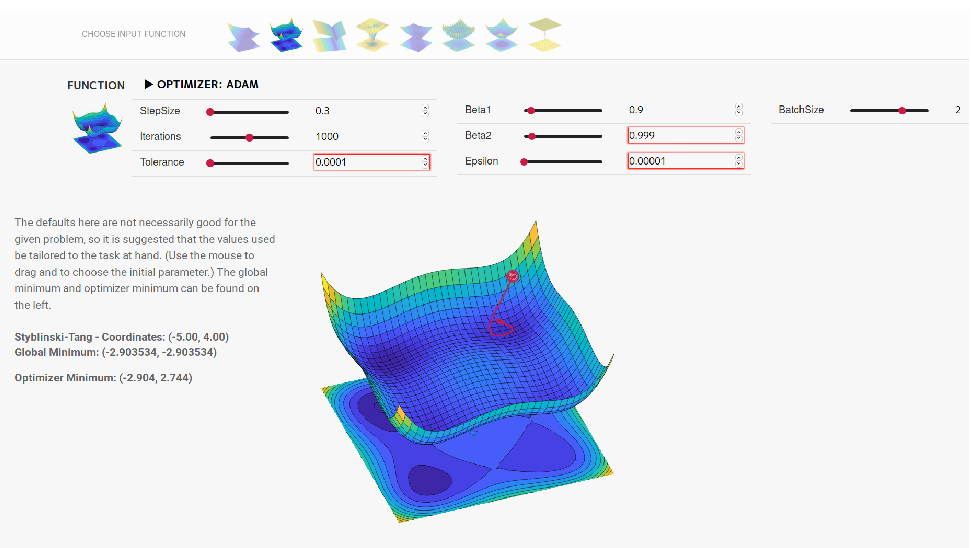}
  \vspace*{-2em}
  \caption{Visualization of the Adam optimizer on the Styblinski-Tang objective
function; this is a screen capture from \url{https://vis.ensmallen.org}.}
  \label{fig:visualization}
\end{figure}

Any other optimizer for separable differentiable objective
functions can be dropped into place in the same manner;
given the large number of available optimizers in {\tt ensmallen},
this support could be used to easily compare optimizers.
In fact, this is exactly done with the interactive
optimizer visualization tool found at \url{https://vis.ensmallen.org}.
Figure~\ref{fig:visualization} shows an example visualization.

\section{Conclusion}
\label{sec:conclusion}

This report introduces and explains {\tt ensmallen}, a C++ mathematical
optimization library that internally uses template metaprogramming to produce efficient
code.  The library is flexible, with support for numerous types of objective
functions, and many implemented optimizers.  It is easy to both implement
objective functions to optimize with {\tt ensmallen}, and to write new
optimizers for inclusion in the library.  {\tt ensmallen} has support for
automatic function inference and callbacks.  Because this is done through
template metaprogramming, there is no additional runtime overhead.
Empirical results show that {\tt ensmallen} outperforms other toolkits 
for similar tasks.

Future work includes the implementation of additional optimizers
and better support for various types of objective functions
(such as better support for constrained functions).
Development is done in an open manner at \mbox{\url{https://github.com/ensmallen/ensmallen}}.
The library uses the permissive BSD license~\cite{Laurent_2008}.
Anyone is welcome to help with the effort and contribute code.

\clearpage
\newpage

\appendix
\titleformat{\section}{\normalfont\Large\bfseries}{Appendix~\thesection:}{1ex}{}
  \section{Inferring Missing Methods -- Details}
\label{sec:automatic_details}

In Section~\ref{sec:automatic}, we described that {\tt ensmallen} is able to
infer missing methods.  For instance, if the user provides a differentiable
objective function that implements {\tt EvaluateWithGradient()}, we can
automatically infer and provide implementations of {\tt Evaluate()} and {\tt
Gradient()} that the optimizer can then use.  (And vice versa: we can infer {\tt
EvaluateWithGradient()} from {\tt Evaluate()} and {\tt Gradient()}.)

Here, we provide details of how this is implemented, focusing specifically on
how {\tt EvaluateWithGradient()} can be inferred, and how this is used in {\tt
ensmallen}'s optimizers.  The same technique is applicable to all of the other
methods that {\tt ensmallen} infers.

\begin{figure}[b!]
\hrule
\vspace{1ex}
\begin{minted}[fontsize=\small]{c++}
// FunctionType is the user-supplied function type to optimize.
// MatType is the user-supplied type of the initial coordinates.
// GradType is the user-specified type of the gradient.
// typename MatType::elem_type represents the internal type held by MatType
//     (e.g., if MatType is `arma::mat`, then this is `double`)
template<typename FunctionType, typename MatType, typename GradType>
typename MatType::elem_type Optimize(FunctionType& function,
                                     MatType& coordinates)
{ 
  // The Function<> mix-in class adds all inferrable methods to `function`
  // So, if `function` has `function.Evaluate()` and `function.Gradient()`, then
  // `fullFunction` will have `fullFunction.EvaluateWithGradient()`.
  typename Function<FunctionType, MatType, GradType> fullFunction(function);
  
  // The rest of the optimizer's code should use `fullFunction`, not `function`.
  ...
}
\end{minted}
\hrule
\vspace*{-0.5em}
\caption{An example implementation of an {\tt ensmallen} optimizer's
{\tt Optimize()} method.
The use of the {\tt Function<>} mix-in class~\cite{smaragdakis2000mixin}
will provide methods that the original {\tt FunctionType} may not have.
}
\label{fig:ex_optimize}
\end{figure}

Using our framework for function inference is straightforward;
an example differentiable optimizer may resemble the code in Figure~\ref{fig:ex_optimize}.
Code for a new optimizer only needs to define a {\tt Function<FunctionType, ...>} wrapper at the
beginning of optimization, and can then expect all three of {\tt Evaluate()},
{\tt Gradient()}, and {\tt EvaluateWithGradient()} to be available.
The class {\tt Function<...>} is like a {\it mix-in} class~\cite{smaragdakis2000mixin};
more accurately, it is a {\it collection} of mix-in classes.

\begin{figure}[b!]
\hrule
\vspace{1ex}
\begin{minted}[fontsize=\small]{c++}
template<typename FunctionType, typename MatType, typename GradType>
class Function :
    public AddEvaluateWithGradient<FunctionType, MatType, GradType>,
    public FunctionType
    ... // many other mixin classes omitted
\end{minted}
\hrule
\vspace*{-0.5em}
\caption{Snippet of the definition of the {\tt Function} class.  The full
implementation contains numerous other mix-in classes like {\tt
AddEvaluateWithGradient}.}
\label{fig:function_snippet}
\end{figure}


Figure~\ref{fig:function_snippet} shows a shortened snippet of the definition
of the {\tt Function} class.
We can see that {\tt Function} inherits methods from the given {\tt
FunctionType} (which is the user-supplied objective function class that is to be
optimized), and also from the {\tt AddEvaluateWithGradient<...>} mixin class.
The key to our approach is that if {\tt FunctionType} class has an {\tt
EvaluateWithGradient()} method, then {\tt AddEvaluateWithGradient<...>} will
provide no functions; if {\tt FunctionType} does {\em not} have an {\tt
EvaluateWithGradient()} method, then {\tt AddEvaluateWithGradient<...>} will
provide an {\tt EvaluateWithGradient()} function.

The details of how this work depend on the SFINAE technique~\cite{abrahams2004c++,Vandevoorde_2018}
and template specialization.  The {\tt AddEvaluateWithGradient} class has five
template parameters, shown in Figure~\ref{fig:aewg}; the last two of these have
default values.

\begin{figure}[b!]
\hrule
\vspace{1ex}
\begin{minted}[fontsize=\small]{c++}
template<typename FunctionType,
         typename MatType,
         typename GradType,
         // Check if FunctionType has at least one non-const Evaluate() or
         // Gradient().
         bool HasEvaluateGradient = ...
         // Check if FunctionType has an EvaluateWithGradient() method already.
         bool HasEvaluateWithGradient = ...
class AddEvaluateWithGradient
{
 public:
  // Provide a dummy overload so the name 'EvaluateWithGradient' exists for this
  // object.
  typename MatType::elem_type EvaluateWithGradient(
      traits::UnconstructableType&);
};
\end{minted}
\hrule
\vspace*{-0.5em}
\caption{Definition of the {\tt AddEvaluateWithGradient} mix-in class.  The
first three template parameters are the `inputs', and the last two template
parameters are computed quantities used for later template specialization.}
\label{fig:aewg}
\end{figure}

For the sake of brevity we omit the expressions for the {\tt bool} parameters
{\tt HasEvaluateGradient} and {\tt HasEvaluateWithGradient}\footnote{Interested
readers can find that code in {\tt
ensmallen\_bits/function/add\_evaluate\_with\_gradient.hpp}.}.  These two
parameters are traits expressions that depend heavily on SFINAE
techniques for method detection; {\tt HasEvaluateGradient} will evaluate to {\tt
true} if {\tt FunctionType} has {\tt Evaluate()} and {\tt Gradient()}; if only
one (or neither) is available, the value is {\tt false}.  {\tt
HasEvaluateWithGradient} will evaluate to {\tt true} if {\tt FunctionType} has
an overload of {\tt EvaluateWithGradient()}, and {\tt false} otherwise.

Using these two boolean template variables, we can then use template
specialization to control the behavior of \linebreak {\tt
AddEvaluateWithGradient} as a function of what is provided by {\tt
FunctionType}.  Specifically, we make two specializations: one for when {\tt
EvaluateWithGradient()} exists, and one for when both {\tt Evaluate()} and {\tt
Gradient()} exist but \linebreak {\tt EvaluateWithGradient()} does not.

\begin{figure}[b!]
\hrule
\vspace{1ex}
\begin{minted}[fontsize=\small]{c++}
template<typename FunctionType,
         typename MatType,
         typename GradType,
         bool HasEvaluateGradient>
class AddEvaluateWithGradient<FunctionType,
                              MatType,
                              GradType,
                              HasEvaluateGradient,
                              true>
{
 public:
  // Reflect the existing EvaluateWithGradient().
  typename MatType::elem_type EvaluateWithGradient(
      const MatType& coordinates, GradType& gradient)
  {
    return static_cast<FunctionType*>(
        static_cast<Function<FunctionType,
                             MatType,
                             GradType>*>(this))->EvaluateWithGradient(
        coordinates, gradient);
  }
};
\end{minted}
\hrule
\vspace*{-0.5em}
\caption{Partial template specialization of {\tt AddEvaluateWithGradient} for
when the given {\tt FunctionType} does have an {\tt EvaluateWithGradient()}
method available already.}
\label{fig:aewg-s1}
\end{figure}

The first specialization, shown in Figure~\ref{fig:aewg-s1}, is for when {\tt
HasEvaluateWithGradient} is {\tt true}---meaning that {\tt FunctionType} already
has {\tt EvaluateWithGradient()}.  In this situation, the implementation of {\tt
AddEvaluateWithGradient::}\linebreak{\tt EvaluateWithGradient()} simply calls
out to {\tt FunctionType::EvaluateWithGradient()}.  However, there is some
complexity here, as to successfully call {\tt
FunctionType::EvaluateWithGradient()}, we must first cast the {\tt this} pointer
to have type {\tt FunctionType}---which can only be done by first casting {\tt
this} to the derived class {\tt Function<...>}.  (This cast is only safe because
we know that we will never create an {\tt AddEvaluateWithGradient} outside of
the context of the {\tt Function<>} class.)

Next, we specialize for the case where {\tt HasEvaluateGradient} is {\tt true}
(i.e., {\tt FunctionType} has both {\tt Evaluate()} and {\tt Gradient()}
methods), and {\tt HasEvaluateWithGradient} is {\tt false} (i.e., there is no
{\tt EvaluateWithGradient()} provided by {\tt FunctionType}).  In this
situation, we intend to synthesize an implementation for {\tt
EvaluateWithGradient()} by using both of the provided {\tt Evaluate()} and {\tt
Gradient()} methods sequentially.  Figure~\ref{fig:aewg-s2} shows this
specialization.

\begin{figure}[b!]
\hrule
\vspace{1ex}
\begin{minted}[fontsize=\small]{c++}
template<typename FunctionType, typename MatType, typename GradType>
class AddEvaluateWithGradient<FunctionType, MatType, GradType, true, false>
{
 public:
  // Use FunctionType's Evaluate() and Gradient().
  typename MatType::elem_type EvaluateWithGradient(const MatType& coordinates,
                                                   GradType& gradient)
  {
    const typename MatType::elem_type objective =
        static_cast<Function<FunctionType,
                             MatType, GradType>*>(this)->Evaluate(coordinates);
    static_cast<Function<FunctionType,
                         MatType,
                         GradType>*>(this)->Gradient(coordinates, gradient);
    return objective;
  }
};
\end{minted}
\hrule
\vspace*{-0.5em}
\caption{Partial template specialization of {\tt AddEvaluateWithGradient} for
when {\tt EvaluateWithGradient()} is not available, but {\tt Evaluate()} and
{\tt Gradient()} are.}
\label{fig:aewg-s2}
\end{figure}

The same complexity with casting is necessary in this specialization too.  The
last specialization is the general case: where there is neither {\tt Evaluate()}
and {\tt Gradient()} nor {\tt EvaluateWithGradient()} provided by {\tt
FunctionType}.  That case is covered by the initial shown implementation of {\tt
AddEvaluateWithGradient}: {\tt EvaluateWithGradient()} is provided with an
argument of type {\tt traits::UnconstructableType}---which is just a class with
a non-public constructor, that cannot be created; and thus, this function cannot
be called.  It is necessary to have this unusable version of {\tt
EvaluateWithGradient()}, though, because the design pattern requires that {\tt
AddEvaluateWithGradient} {\em always} provides a method with the name {\tt
EvaluateWithGradient()}.

This does, however, mean that users can get long and confusing error messages if
the optimizer attempts to instantiate the overload of {\tt
EvaluateWithGradient()} with {\tt traits::UnconstructableType}.  But, remember
that this situation can only happen when using a differentiable optimizer if
the user provided a {\tt FunctionType} that {\it (a)} does not have both {\tt
Evaluate()} or {\tt Gradient()}, or {\it (b)} does not have {\tt
EvaluateWithGradient()}.  This is something that we already have code to
detect---that is the code that computes the values of the boolean template
parameters {\tt HasEvaluateGradient} and {\tt HasEvaluateWithGradient}.
Therefore, we simply use a {\tt static\_assert()} to provide a clear and
understandable error message at compile time when neither of those values are
{\tt true}.  We encapsulate this in a convenient function that can be added at
the beginning of the {\tt Optimize()} method:

\begin{minted}[fontsize=\small]{c++}
  CheckFunctionTypeAPI<FunctionType, MatType, GradType>();
\end{minted}

This technique is adapted to each objective function type that {\tt ensmallen}
supports, in order to provide straightforward error messages of the form shown in Figure~\ref{fig:example_error_message_1}.
This {\tt static\_assert()} failure output appears typically at the end of the
error output, which is far preferable to the output without {\tt
CheckFunctionTypeAPI}.  Example output produced by {\tt clang++} is shown in
Figure~\ref{fig:example_error_message_2}.

\begin{figure}[!tb]
\hrule
\vspace{1ex}
\begin{minted}[fontsize=\small]{text}
    ...
    include/ensmallen_bits/function/static_checks.hpp:292:3: error: static_assert
          failed due to requirement
          'CheckEvaluateWithGradient<ens::Function<TestFunction, arma::Mat<double>,
          arma::Mat<double> >, arma::Mat<double>, arma::Mat<double> >::value' "The
          FunctionType does not have a correct definition of EvaluateWithGradient().
          Please check that the FunctionType fully satisfies the requirements of the
          FunctionType API; see the optimizer tutorial for more details."
      static_assert(
      ^
    ...
\end{minted}
\hrule
\caption{Example error output produced by {\tt clang++} when the given objective
function is missing methods and {\tt static\_assert()}s are used.  Compare with
Figure~\ref{fig:example_error_message_2}, which is far less clear.}
\label{fig:example_error_message_1}
\end{figure}

\begin{figure}[!tb]
\hrule
\vspace{1ex}
\begin{minted}[fontsize=\small]{text}
    ...
    include/ensmallen_bits/lbfgs/lbfgs_impl.hpp:376:30: error: no matching member
          function for call to 'EvaluateWithGradient'
      ElemType functionValue = f.EvaluateWithGradient(iterate, gradient);
                               ~~^~~~~~~~~~~~~~~~~~~~
    include/ensmallen_bits/lbfgs/lbfgs.hpp:97:12: note: in instantiation of function
          template specialization 'ens::L_BFGS::Optimize<TestFunction,
          arma::Mat<double>, arma::Mat<double>>' requested here
        return Optimize<SeparableFunctionType, MatType, MatType,
               ^
    test.cpp:19:31: note: in instantiation of function template specialization
          'ens::L_BFGS::Optimize<TestFunction, arma::Mat<double>>' requested here
      const double result = lbfgs.Optimize(f, parameters);
                                  ^
    include/ensmallen_bits/function/add_evaluate_with_gradient.hpp:225:38: note:~
          candidate function not viable: requires 1 argument, but 2 were provided
      static typename MatType::elem_type EvaluateWithGradient(
    ...
\end{minted}
\hrule
\caption{Example error output produced by {\tt clang++} when the given objective
function is missing methods, and {\tt static\_assert()}s are not used.  Compare
with Figure~\ref{fig:example_error_message_1}, which is much clearer.}
\label{fig:example_error_message_2}
\end{figure}

Note that the error message above can be highly misleading: it's not actually a
requirement that {\tt EvaluateWithGradient()} be supplied by the objective
function class!  That can be automatically inferred, just like discussed above,
but only if {\tt Evaluate()} and {\tt Gradient()} are both available.  In the
case of the example code that generated the errors above, {\tt Gradient()} was
not implemented---thus, the given error message is actually somewhat inaccurate
and confusing!

The last piece of the puzzle is making sure that when {\tt
Function<...>::EvaluateWithGradient()} is called, that this always calls {\tt
AddEvaluateWithGradient::EvaluateWithGradient()}.  This can be done by a simple
{\tt using} declaration in the body of the {\tt Function} class:

\begin{minted}[fontsize=\small]{c++}
    using AddEvaluateWithGradient<FunctionType,
                                  MatType,
                                  GradType>::EvaluateWithGradient;
\end{minted}

This is all that is needed to infer and synthesize an {\tt
EvaluateWithGradient()} method when the user provided an objective function that
only has {\tt Evaluate()} and {\tt Gradient()}.  There is one detail we have
omitted, though---the code that we have shown here handles non-{\tt const}
and non-{\tt static} versions of methods provided by the user.  Separate
auxiliary structures like {\tt AddEvaluateWithGradientConst}
and {\tt AddEvaluateWithGradientStatic}
are used for function inference in those cases; the general design is identical.

All of these pieces put together result in a clean interface that inference of
methods that users did not provide in their objective function.  Further, this
all happens at {\it compile time} and thus there is no runtime penalty.
Auxiliary structures like {\tt AddEvaluateWithGradient} should be optimized out
by the compiler.

All of the code involved with automatic generation of missing methods can be
found in the {\tt ensmallen} source code in the directory {\tt
include/ensmallen\_bits/function/}.

  \section{Callbacks -- Details}
\label{sec:callback_details}

In Section~\ref{sec:callbacks}, we introduced {\tt ensmallen}'s support for
callbacks.  This support is implemented via templates---like much of the rest of
{\tt ensmallen}.  This design strategy was chosen in order to minimize runtime
overhead caused by callbacks, and produce machine code roughly equivalent to the
machine code that would have been produced if the callback had been directly
integrated into the optimizer's code.

In this section, we detail the implementation of these callbacks and show how to
implement optimizers that are capable of handling these callbacks.

An optimizer that supports callbacks should call out to the static methods in
the {\tt Callback} class inside of its {\tt Optimize()} method, as below:
\begin{minted}[fontsize=\small]{c++}
    double functionValue = function.Evaluate(coordinates);
    bool   terminate     = Callback::Evaluate(*this, function, coordinates, functionValue, callbacks...);
\end{minted}

In the above snippet, {\tt function} represents the function being optimized,
{\tt *this} is the optimizer itself, {\tt coordinates} represents the current
coordinates of the optimization, and {\tt callbacks} is a template vararg pack
containing all of the given callbacks.
Note that this means {\tt callbacks} can be empty.

The {\tt Callback} class contains a top-level method to use for each of the
callback types supported by {\tt ensmallen}, as listed in Table~\ref{tab:callback_list}.
There is a {\tt Callback::Gradient()} method, {\tt Callback::EvaluateConstraint()},
{\tt Callback::BeginEpoch()}, and so forth. 

\begin{figure}[b!]
\hrule
\vspace{1ex}
\begin{minted}[fontsize=\small]{c++}
template<typename OptimizerType, typename FunctionType, typename MatType, typename... CallbackTypes>
static bool Callback::Evaluate(OptimizerType& optimizer,
                               FunctionType& function,
                               const MatType& coordinates,
                               const double objective,
                               CallbackTypes&... callbacks)
{
  // This will return immediately once a callback returns true.
  bool result = false;
  (void) std::initializer_list<bool>{ result =
      result || Callback::EvaluateFunction(callbacks, optimizer, function,
      coordinates, objective)... };
   return result;
}
\end{minted}
\hrule
\vspace*{-0.5em}
\caption{Implementation of {\tt Callback::Evaluate()},
showing part of the process of calling each callback
given in the template vararg pack {\tt callbacks}.
}
\label{fig:callback_evaluate}
\end{figure}

The {\tt Callback::Evaluate()} method calls the {\tt Evaluate()} method of every
given callback that has an {\tt Evaluate()} method implemented.  This means that
there is some amount of difficulty we have to handle: not every callback in {\tt
callbacks...} will have an {\tt Evaluate()} method available.  Thus, we can use
the same techniques as in Section~\ref{sec:automatic} to detect, via SFINAE,
whether an {\tt Evaluate()} method exists for a given callback, and then perform
the correct action based on the result.

The definition of {\tt Callback::Evaluate()} is given in
Figure~\ref{fig:callback_evaluate}.  Each callback function takes several
template parameters, including {\tt OptimizerType} (the type of optimizer that
is being used), {\tt FunctionType} (the type of function being optimized), {\tt
MatType} (the matrix type used to store the coordinates), and {\tt
CallbackTypes} (the set of types of callbacks).  Of these, {\tt CallbackTypes}
is the most important.  Since it is a template vararg, the
type {\tt CallbackTypes} is actually a pack that corresponds to every type of
every callback that must be called.

The implementation of the function unpacks the {\tt callbacks},
calling each callback's {\tt Evaluate()} method (if it exists) in sequence,
and terminating early if any of these callbacks returns {\tt true}.
Each callback's {\tt Evaluate()} method is called through the helper
function {\tt Callback::EvaluateFunction()},
which uses SFINAE traits to control behavior depending on whether
the given callback has an {\tt Evaluate()} method or not.
Figure~\ref{fig:callback_evaluate_function} shows the two overloads of
{\tt Callback::EvaluateFunction()}.

\begin{figure}[t!]
\hrule
\vspace{1ex}
\begin{minted}[fontsize=\small]{c++}
template<typename CallbackType, typename OptimizerType, typename FunctionType, typename MatType>
static typename std::enable_if<callbacks::traits::HasEvaluateSignature<
    CallbackType, OptimizerType, FunctionType, MatType>::value, bool>::type
EvaluateFunction(CallbackType& callback,
                 OptimizerType& optimizer,
                 FunctionType& function,
                 const MatType& coordinates,
                 const double objective)
{
  return (const_cast<CallbackType&>(callback).Evaluate(optimizer, function, coordinates, objective),
      false);
}

template<typename CallbackType, typename OptimizerType, typename FunctionType, typename MatType>
static typename std::enable_if<!callbacks::traits::HasEvaluateSignature<
    CallbackType, OptimizerType, FunctionType, MatType>::value, bool>::type
EvaluateFunction(CallbackType& /* callback */,
                 OptimizerType& /* optimizer */,
                 FunctionType& /* function */,
                 const MatType& /* coordinates */,
                 const double /* objective */)
{ return false; }
\end{minted}
\hrule
\vspace*{-0.5em}
\label{fig:callback_evaluate_function}
\caption{Implementation of {\tt Callback::EvaluateFunction()}.  Two overloads
are specified: the first is used when {\tt CallbackType} has an {\tt Evaluate()}
method, and the second is used when {\tt CallbackType} does not have an {\tt
Evaluate()} method.  SFINAE is used, via {\tt
callbacks::traits::HasEvaluateSignature}, to determine which overload to use.}
\end{figure}

Similar to Section~\ref{sec:automatic}, a traits class is used to determine
which of the two overloads of {\tt EvaluateFunction()} to call.  The traits
class {\tt callbacks::traits::HasEvaluateSignature<...>::value}
evaluates to {\tt true} only when the inner callback and arguments can form a
valid {\tt Evaluate()} signature.  Thus, the correct overload is called when
{\tt EvaluateFunction()} is called from {\tt Callback::Evaluate()}.

Each of the other callbacks in the {\tt Callback} class is implemented in
virtually identical form, although with different arguments depending on what
the callback is.  For further details on the callback infrastructure,
examples are provided in the code repository in the directory
{\tt include/ensmallen\_bits/callbacks/}.

  \section{Static Assertions to Check Matrix Traits}
\label{sec:templated_optimize_details}

In Section~\ref{sec:templated_optimize}, we showed that {\tt ensmallen} uses
templates to allow arbitrary types to be used for the matrix type or gradient
type.
However, care must be taken there to avoid a drawback with C++ template metaprogramming,
which is the issue of compiler errors.
In Figure~\ref{fig:gd}, the compiler may fail to substitute a given {\tt MatType}
into the code for {\tt GradientDescent::Optimize()}. A~typical reason might be that
a required method of {\tt MatType} or {\tt FunctionType} is not available.
For example, the given {\tt MatType} does not have an {\tt operator\*()} method.
Ordinarily, current C++ compilers typically emit a long list of error messages,
with unnecessarily detailed and distracting information.
The internal framework in {\tt ensmallen} avoids this issue via the use of
C++ compile-time static assertions ({\tt static\_assert()}),
resulting in considerably clearer error messages.

The framework provides a set of utility methods for checking traits of {\tt
MatType} and {\tt GradType} (for differentiable objective functions) inside of
an optimizer's {\tt Optimize()} method:

\begin{itemize}
  \item {\tt RequireFloatingPointType<MatType>()}: this requires that the
element type held by {\tt MatType} is either {\tt float} or {\tt double}.
This is generally needed by optimizers that use Armadillo functionality which
ends up calling LAPACK functions~\cite{anderson1999lapack}.

  \item {\tt RequireSameInternalTypes<Mat1Type, Mat2Type>()}: this requires that
{\tt Mat1Type} and {\tt Mat2Type} use the same type to hold elements.  This is
useful for differentiable optimizers, where the user is allowed to specify
different objective matrix types ({\tt MatType}) and gradient matrix types ({\tt
GradType}).  This function allows us to require that the same type (e.g., {\tt
int} or {\tt float} or {\tt double}) is used for both {\tt MatType} and {\tt
GradType}.

  \item {\tt RequireDenseFloatingPointType<MatType>()}: this requires that the
element type held by {\tt MatType} is either {\tt float} or {\tt double}, and
that the representation used for storage by {\tt MatType} is dense.
\end{itemize}

These methods would typically be called at the top of the implementation of an
optimizer's {\tt Optimize()} method.

At the time of writing, these are the only three checks that have
been needed, but it is easy to add more.  The implementation of these functions
is just a {\tt static\_assert()} that uses some underlying traits of the given
template parameters.  For example, Figure~\ref{fig:rsit} is the implementation
of {\tt RequireSameInternalTypes<...>()}.

Note that users who would like to continue despite these {\tt static\_assert()}
checks failing may simply define the macro {\tt ENS\_DISABLE\_TYPE\_CHECKS} in
their code before the line {\tt \#include <ensmallen.hpp>}.

\begin{figure}[t]
\hrule
\vspace{1ex}
\begin{minted}[fontsize=\small]{c++}
/**
 * Require that the internal element type of the matrix type and gradient type
 * are the same.  A static_assert() will fail if not.
 */
template<typename MatType, typename GradType>
void RequireSameInternalTypes()
{
#ifndef ENS_DISABLE_TYPE_CHECKS
  static_assert(std::is_same<typename MatType::elem_type,
                             typename GradType::elem_type>::value,
      "The internal element types of the given MatType and GradType must be "
      "identical, or it is not known to work!  If you would like to try "
      "anyway, set the preprocessor macro ENS_DISABLE_TYPE_CHECKS before "
      "including ensmallen.hpp.  However, you get to pick up all the pieces if "
      "there is a failure!");
#endif
}
\end{minted}
\hrule
\vspace*{-0.5em}
\caption{Implementation of {\tt RequireSameInternalTypes()} from internal
{\tt ensmallen} code.
}
\label{fig:rsit}
\end{figure}


\clearpage
\newpage
\bibliographystyle{ieee}
\bibliography{refs}

\end{document}